\documentclass[runningheads]{llncs}

\usepackage{cite}

\usepackage[T1]{fontenc}

\usepackage[hyphens]{url}

\usepackage[colorlinks=true,linkcolor=blue,citecolor=blue,urlcolor=blue]{hyperref}

\usepackage{graphicx}

\usepackage{color}
\usepackage{amsmath}
\usepackage{subcaption}

\iftrue
  \newcommand\comment[1]{\textbf{\color{blue}[TODO: #1]}}
  \newcommand\reviewfix[1]{\textbf{\color{red}[#1]}}
\else
  \newcommand\comment[1]{\EatSpacesHack}
  \newcommand\reviewfix[1]{\EatSpacesHack}
\fi

\makeatletter
\renewcommand\section{\@startsection{section}{1}{\z@}%
                       {-8\p@ \@plus -4\p@ \@minus -4\p@}%
                       {6\p@ \@plus 4\p@ \@minus 4\p@}%
                       {\normalfont\large\bfseries\boldmath
                        \rightskip=\z@ \@plus 8em\pretolerance=10000 }}
\renewcommand\subsection{\@startsection{subsection}{2}{\z@}%
                       {-8\p@ \@plus -4\p@ \@minus -4\p@}%
                       {6\p@ \@plus 4\p@ \@minus 4\p@}%
                       {\normalfont\normalsize\bfseries\boldmath
                        \rightskip=\z@ \@plus 8em\pretolerance=10000 }}
\renewcommand\subsubsection{\@startsection{subsubsection}{3}{\z@}%
                       {-4\p@ \@plus -4\p@ \@minus -4\p@}%
                       {-1.5em \@plus -0.22em \@minus -0.1em}%
                       {\normalfont\normalsize\bfseries\boldmath}}
\makeatother

\begin{document}
\title{DARSAN: A Decentralized Review System Suitable for NFT Marketplaces}

%
%

\author{Sulyab Thottungal Valapu\inst{1} \and
    Tamoghna Sarkar\inst{1} \and
    Jared Coleman\inst{1} \and \\
    Anusha Avyukt\inst{1} \and
    Hugo Embrechts\inst{2} \and
    Dimitri Torfs\inst{2} \and
    Michele Minelli\inst{2} \and
    Bhaskar Krishnamachari\inst{1}}

\authorrunning{S. Thottungal Valapu et al.}
%

\institute{University of Southern California, Los Angeles, USA
    \email{\{thottung,tsarkar,jaredcol,aavyukt,bkrishna\}@usc.edu}\\
    \and
    SONY R\&D Center Brussels Laboratory, Brussels, Belgium\\
    \email{\{hugo.embrechts,dimitri.torfs,michele.minelli\}@sony.com}}

\maketitle              

\begin{abstract}
    We introduce DARSAN, a decentralized review system designed for Non-Fungible Token (NFT) marketplaces, to address the challenge of verifying the quality of highly resalable products with few verified buyers by incentivizing unbiased reviews.
    DARSAN works by iteratively selecting a group of reviewers (called ``experts'') who are likely to both accurately predict the objective popularity and assess some subjective quality of the assets uniquely associated with NFTs.
    The system consists of a two-phased review process: a ``pre-listing'' phase where only experts can review the product, and a ``pre-sale'' phase where any reviewer on the system can review the product.
    Upon completion of the sale, DARSAN distributes incentives to the participants and selects the next generation of experts based on the performance of both experts and non-expert reviewers.
    We evaluate DARSAN through simulation and show that, once bootstrapped with an initial set of appropriately chosen experts, DARSAN favors honest reviewers and improves the quality of the expert pool over time without any external intervention even in the presence of potentially malicious participants.

\end{abstract}

\section{Introduction}
Ratings and reviews have a significant impact on the perception of potential customers regarding the quality of a product~\cite{hu_online_2008,zhu_impact_2010}.
Therefore, it is in the interest of online marketplaces to promote helpful and high-quality reviews, while demoting biased or low-value ones.
Although challenging for any online marketplace~\cite{DBLP:conf/icbc2/AvyuktRK21}, designing a review system for Non-Fungible Token (NFT) marketplaces presents unique difficulties due to their scarcity and high resale potential.
The sale of an NFT collection, typically limited to a few hundred or thousand pieces, creates two groups of users: a minority who own one of the NFTs and a supermajority who do not.
The high resale potential of NFTs creates an incentive for the minority to rate them highly, irrespective of their true opinion, in the hopes of reselling them at a higher price in the future.
Conversely, the non-owning supermajority have less incentive to rate the NFTs positively and may even rate them poorly to decrease their value, thereby increasing their chances of obtaining the NFTs for a lower price in the future, or to increase the relative value of the NFTs they own.
As of February 2023, none of the top five NFT marketplaces ranked by trading volume (Blur, OpenSea, X2Y2, Magic Eden, LooksRare)~\cite{coingecko_most_2023} have integrated rating or review systems, which further supports this argument.
Instead, these marketplaces rely on indirect metrics such as the number of users who viewed or ``favorited'' an NFT, which can be easily gamed.

\begin{figure}[hbt]
    \centering
    \includegraphics[scale=0.8]{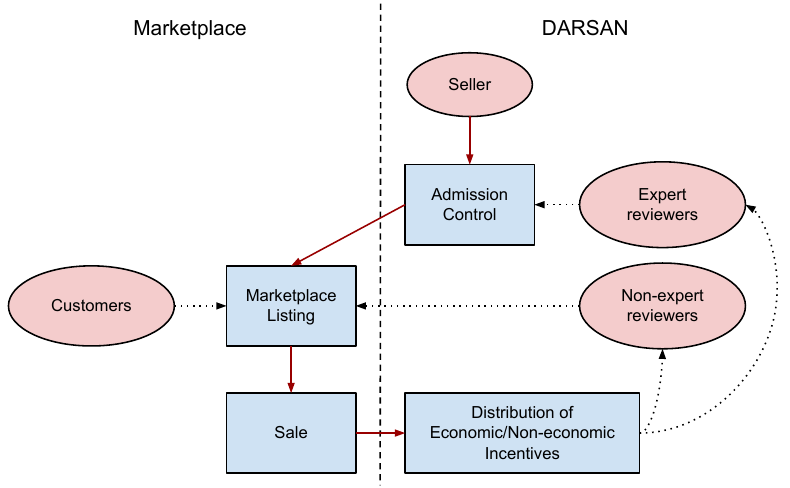}
    \caption{
        Illustration of the integration of DARSAN within an NFT marketplace.
        The red ovals symbolize different roles, i.e, an individual may simultaneously act as a customer and a reviewer.
    }
    \label{overall}
\end{figure}

In this paper, we introduce DARSAN, a decentralized review system designed for Non-Fungible Token (NFT) marketplaces that aims to address this issue.
DARSAN utilizes an approach where a group of reviewers, known as ``experts'' are iteratively selected based on their ability to accurately predict the objective popularity and assess some subjective quality of the assets uniquely associated with NFTs.
While the objective popularity is measured using ground truths associated with sales, such as sale price or volume, DARSAN does not require the system to explicitly define \emph{any} rubric to assess the subjective quality of an asset.
Instead, it relies on expert consensus to implicitly establish the rubric at any given time.
The review process consists of two phases: a ``pre-listing'' phase, where only experts can review the product, and a ``pre-sale'' phase, where any reviewer on the system can review the product.
After the sale, DARSAN distributes economic as well as non-economic incentives to the participants and selects the next generation of experts based on the performance of both expert and non-expert reviewers.
Fig.~\ref{overall} illustrates the interaction between DARSAN and an NFT marketplace.

DARSAN has numerous features that make it an attractive option to marketplace owners and users alike.
One of its most significant advantages is \emph{versatility}, as it enables marketplace owners to achieve a balance between the relative significance of subjective opinions (e.g., expert reviews) and objective data (e.g., sales metrics) by choosing system parameters appropriately.
Once deployed, DARSAN is \emph{self-sufficient}, requiring no further involvement from the marketplace owner.
Furthermore, DARSAN ensures \emph{transparency} through blockchain logging, which allows for all actions to be publicly verifiable.
This eliminates the possibility of manipulative practices by marketplace owners, such as ``shadow banning'' and falsification of product ratings, which have been increasingly reported in recent years~\cite{kalra_special_2021, soper_amazon_2018, le_merrer_setting_2021}.

We performed numerical simulations to assess the effectiveness of DARSAN in identifying new generations of experts and its resistance to adversarial behavior from expert as well as non-expert participants.
Our findings suggest that once the system is bootstrapped with an initial set of ``appropriately'' chosen experts, DARSAN incentivizes honest reviewers, leading to an improvement in the quality of the expert pool over time, even in the presence of potentially malicious participants.
The system accomplishes this without any external intervention, which makes it an ideal option for a decentralized review system for highly resalable products that have few verified buyers, such as NFT marketplaces and art markets.
Moreover, the ability of the system to combine subjective opinions and objective metrics for decentralized decision-making makes it ideal for integration into other decentralized systems.

The paper is structured as follows.
Section~\ref{sec:relatedwork} presents an overview of existing literature on the topic and assesses the strengths and limitations of current review systems in the context of NFT marketplaces.
In Section~\ref{sec:architecture}, the architecture of DARSAN is described in detail, including its typical workflow.
Section~\ref{sec:evaluation} analyses the proposed system using numerical simulations.
Finally, we conclude the paper in Section~\ref{sec:conclusion}.

\section{Related Work} \label{sec:relatedwork}
The economics of collectibles and art markets have been studied extensively over the past several decades~\cite{stoller_economics_1984, frey_art_1997, agnello_investment_2002}.
In the recent years, efforts have been made to study the economics of NFTs from various perspectives such as pricing~\cite{dowling_is_2022}, returns~\cite{umar_covid-19_2022} and investment risk~\cite{karim_examining_2022}.
Despite these efforts, the role of ratings and reviews in art, collectible, or NFT marketplaces remains under-explored.

A number of recently proposed blockchain platforms~\cite{DBLP:conf/bwcca/JavaidZAKNJ19,DBLP:journals/csur/HasanBB23} use reputation systems to properly incentivize correct behavior by the platform users.
Steemit, a blogging and social media platform, has its own tokens which are used to incentivize users to post quality content~\cite{steem2017steem}.
Relevant, a news-sharing and discussion platform, introduces the concept of ``reputation contexts'' which allow users to earn reputation for different categories of content (e.g. politics, sports, technology, etc.) which allows users to specialize in and earn reputation for their expertise in specific categories~\cite{balasanov2018technical}.
In Steemit, curators (users who upvote content) are rewarded with tokens based on the performance of the content they upvote~\cite{steem2017steem}.
In Relevant, users can predict the performance of content they upvote and earn tokens based on their predictions.
A few academic papers have also proposed blockchain-based review systems.
A reputation-based system for IoT marketplaces has been proposed where device owners gain reputation when data consumers use their data and leave positive reviews~\cite{DBLP:conf/bwcca/JavaidZAKNJ19}.

On the academic side, a few rating and review systems have been proposed, including one which uses control products with known quality to randomly test reviewer honesty/ability~\cite{DBLP:conf/icbc2/AvyuktRK21}.
Implementing this kind of ``mystery shopper'' approach for NFT marketplaces, though, is challenging because it's difficult to introduce a control product with a determined quality when the quality in question is subjective.
It has also been suggested that user ratings could potentially enhance the transparency and trustworthiness of data in a decentralized data marketplace for smart cities~\cite{DBLP:conf/isc2/RamachandranRK18}.
We believe the in-depth analysis in this paper helps support this claim.
Most similar to our work, one solution leverages Lina.Review~\cite{linareview}, a blockchain-based review system, to implement a reputation system with two classes of users: \textit{Helpers}, who are paid for high-quality reviews and \textit{Common Users}, who are promoted to Helpers based on their high-quality reviews~\cite{DBLP:journals/tcss/GlenskiPW17}.
This solution, however, relies solely on likes to determine review quality (rather than sale price or some other ground truth metric, as our proposed solution does) and lacks a satisfying incentive analysis.
ReviewChain\cite{DBLP:conf/ithings/WangZK18}, a decentralized blockchain-based review system, ensures the authenticity and integrity of reviews by maintaining singular identities for reviewers and confirming product purchase by reviewers, while our study focuses on different aspects - namely quality of reviews and product ranking.

To the best of our knowledge, our solution is the first decentralized rating and review system for NFT marketplaces that incentivizes high-quality sellers to use the platform and reviewers to provide unbiased and high-quality reviews.

\section{Proposed Architecture} \label{sec:architecture}
\subsection{Use of Blockchain}
Although our proposed architecture is blockchain-based, any transparent, publicly auditable, and immutable ledger with smart contract-like capabilities is suitable for our purpose.
Governance decisions regarding the choice of consensus mechanism, participants in the consensus process, and related matters are entirely at the discretion of system designers.

\subsection{Roles and Concepts}
The entity that owns the NFT marketplace and the associated review system is referred to as the \emph{authority}.
Prior to deploying the system, the authority selects a set of \emph{areas of expertise} that are relevant to the marketplace's offerings.
For example, if the marketplace specializes in gaming-related NFTs, the areas of expertise may include art, music, first-person shooter (FPS) games, etc.
Entities who list NFTs for sale on the marketplace are known as \emph{sellers} and can include individual artists and/or authorized agents working on behalf of artists.

At the core of the review system are \emph{reviewers}, who are responsible for evaluating the products listed on the marketplace as well as endorsing/reporting other reviewers.
Through these actions, reviewers earn \emph{expertise} points in the area(s) of expertise relevant to their actions.
Expertise is a non-negative numerical value that quantifies the value a reviewer's opinions have in a particular area of expertise.
The higher a reviewer's expertise score, the more impact their choices have on the system.
At any given time $t$, the top $k$ reviewers with the highest expertise in a particular area are considered \emph{experts} of that area.
Thus, once the system is deployed, participants may enter or exit the pool of experts over time.

\subsection{Pre-Deployment (Off-Chain) Phase}
Prior to deploying the system, the authority selects the areas of expertise as well as the initial set of experts in each of those areas.
The initial set of experts begin with a pre-determined high expertise score in their area of expertise whereas all other reviewers join the system with zero expertise in all areas.
It is important to emphasize that the authority's involvement is confined to the pre-deployment phase.
After the completion of this phase, no further actions are required from the authority to maintain the system.

\subsection{Post-Deployment (On-Chain) Phase} \label{workflow}
The post-deployment phase is considered per-asset, and involves the entire life cycle of an asset on the marketplace including admission control, marketplace listing, and sale.
We refer to the entire life cycle of an asset as one \textit{round}.
Each round comprises of a total of eleven steps that can be categorized into \emph{pre-listing}, \emph{pre-sale} and \emph{post-sale}.
We now describe each step in detail.

\subsubsection{Pre-Listing}
The pre-listing period, exclusive to experts in the relevant areas, focuses on admission control, i.e, determining whether the asset should be eligible for listing on the marketplace.
The various steps involved in the pre-listing period are depicted in Figure~\ref{fig:prelisting}.
It consists of the following steps:

\begin{figure}[htb]
    \centering
    \includegraphics{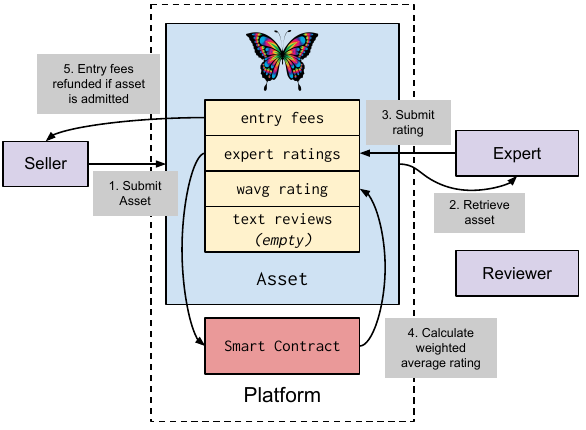}
    \caption{
        Steps involved in the pre-listing period.
    }
    \label{fig:prelisting}
\end{figure}

\paragraph{Step 1.}
The process begins when a seller, who has been approved by the authority, submits an asset to the marketplace.
As our primary focus is on the review system, the specific details of the seller approval process are left to the authority.
In addition to the asset itself, the seller indicates one or more \emph{area tags} that are applicable to the asset and also stakes an \emph{entry fee} stipulated by the marketplace.
These area tags are used to identify the relevant set of experts responsible for evaluating the asset.
The entry fee serves as a deterrent against spam and is forfeited if the asset fails to pass the admission control.

\paragraph{Step 2.}
Once the seller submits the asset, it is assigned to all the experts in the corresponding areas based on the specified area tags.

\paragraph{Step 3.}
Experts are given a predetermined amount of time to submit a numerical rating of the asset, say, on a scale from 0 to 5.
The rating provided by expert $i$ is denoted as $r_i$.
Experts also have the option to submit text reviews in addition to the rating, allowing them to express their opinions about the asset in detail.
Experts may opt to commit to a rating utilizing cryptographic techniques, withholding disclosure of the rating they committed to until the expiration of the designated time window.

\paragraph{Step 4.}
Upon the expiration of the designated time window, the smart contract calculates the weighted average numerical rating, denoted as $\bar{r}$, using the expertise scores of each expert as the respective weights.

\paragraph{Step 5.}
The smart contract makes the decision regarding the admission of the asset to the marketplace by comparing $\bar{r}$ with a minimum rating threshold ($thresh$) established by the authority.
The asset is admitted if $\bar{r} \geq thresh$, and rejected otherwise.
In the case of rejection, the entry fee staked by the seller is forfeited and added to the economic incentive pool, and the round ends.
The subsequent steps are only executed if the asset is successfully admitted to the marketplace.

\subsubsection{Pre-Sale}
During the pre-sale period, all non-expert reviewers are provided with the chance to review the asset and optionally endorse reviews contributed by other reviewers.
Fig.~\ref{fig:presale} depicts the steps involved.
We now describe each step in detail:

\begin{figure}[htb]
    \centering
    \includegraphics{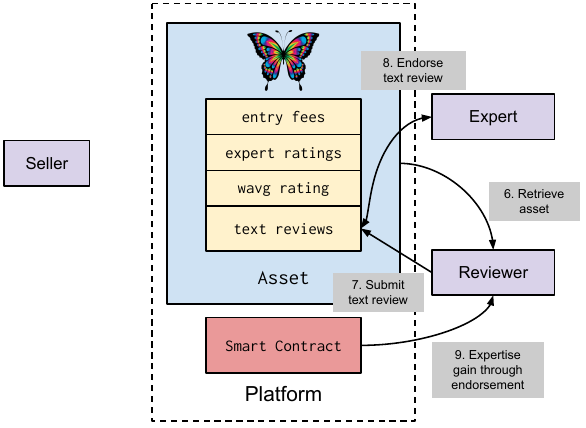}
    \caption{
        Steps involved in the pre-sale period.
    }
    \label{fig:presale}
\end{figure}

\paragraph{Step 6.}
The admitted asset is displayed under the sales listing in decreasing order of $\bar{r}$, i.e, if asset $a_1$ has a higher weighted average rating than asset $a_2$, then $a_1$ is listed first, followed by $a_2$.
Once the asset has been listed on the marketplace, it becomes accessible to all participants, including potential reviewers.

\paragraph{Step 7.}
Reviewers have the choice to submit a text review expressing their personal opinions about the asset, and/or make predictions about its relative sales performance (i.e, its objective popularity).
The exact method of providing the relative sales performance prediction is left to the marketplace.
For example, one viable method could involve reviewers ranking the currently listed assets based on their predicted popular demand from least to most popular.
For resilience against Sybil attacks, the reviewers are also required to stake a nominal amount (chosen by the marketplace) while submitting reviews and/or predictions, that will be refunded after a stipulated time unless spam activity has been detected.

\paragraph{Step 8.}
Once a reviewer submits their review of the asset, experts corresponding to the relevant area(s) can retrieve the review, and optionally \emph{endorse} it as described in the next step.

\paragraph{Step 9.} \label{arch_endorsement}
Optionally, reviewers may \emph{endorse} text reviews by other reviewers.
All reviewers, regardless of their expertise level, are allocated a stipulated amount of \emph{endorsement power} per asset, as determined by the authority.
In this paper, we will consider the model where each reviewer is granted exactly one endorsement to utilize per asset that will expire if left unused.
\footnote[1]{Other models of endorsement power are possible, which could result in different endorsement strategies.}
Thus, reviewers have the option to endorse a single text review (excluding their own) for each asset.
Endorsements affect the way reviewers gain expertise in two ways:

\emph{1. Expertise gain by the endorsee.}
While all reviewers have the ability to make endorsements, endorsements from experts result in the endorsee gaining some expertise in the corresponding area(s).
The amount of expertise gained by the endorsee is determined by two factors: the expertise score of the endorsing expert, and the difference in expertise between the expert and the endorsee.
Mathematically, the expertise gained by a reviewer $r$ due to an endorsement from expert $e$, denoted as $\Delta(r,e)$, can be represented as
\begin{equation}
    \Delta(r,e)=mingain(exp_e)+addgain(max(0, exp_e-exp_r))
\end{equation}
where $exp_i$ is the expertise of person $i$, and $mingain()$ and $addgain()$ are functions defined by the authority that calculate the minimum expertise gain and the additional expertise gain respectively.
Therefore, the minimum expertise gain is determined by the expertise score of the endorsing expert, and any additional expertise gain depends on the difference in expertise score between the endorsing expert and the endorsee.

\emph{2. Expertise gain by investors.}
Endorosements act as \emph{investments} made by the endorser in the endorsee.
By making this investment, the endorser establishes a stake and gains a proportionate fraction of the expertise acquired by the endorsee in subsequent rounds.
The introduction of investments within the system is intended to incentivize experts to identify and endorse reviewers who are likely to consistently perform well over time, thus helping the system in identifying high-quality reviewers who eventually may progress to become experts themselves.
Since an endorsement by an expert results in an expertise gain for the endorsee, all investors that invested in the endorsee up until the previous round will gain expertise proportional to their ``share'' of investment.
However, to prevent gaming of the system, you cannot gain investment dividends from your own subsequent endorsements.
Mathematically, the expertise gain by investor $i$ due to a reviewer $r$ being endorsed by an expert $e$ can be represented as
\begin{equation}
    dividend(i,r,e)=\frac{c_1\times\Delta(r,e)\times invshare^i_r}{\sum_{j\in R} invshare^j_r}
\end{equation}
where $c_1$ is a positive constant determined by the authority, $invshare^i_r$ is the number of times investor $i$ has endorsed reviewer $r$, and $R$ is the set of all reviewers.

\subsubsection{Post-Sale}
Following the completion of the sale, the post-sale computations are performed, including the distribution of expertise points based on observed sales metrics, as well as the selection of the ``next generation'' of experts for the subsequent round.
The post-sale period, depicted in Fig.~\ref{fig:postsale}, comprises of the following steps:

\begin{figure}[htb]
    \centering
    \includegraphics{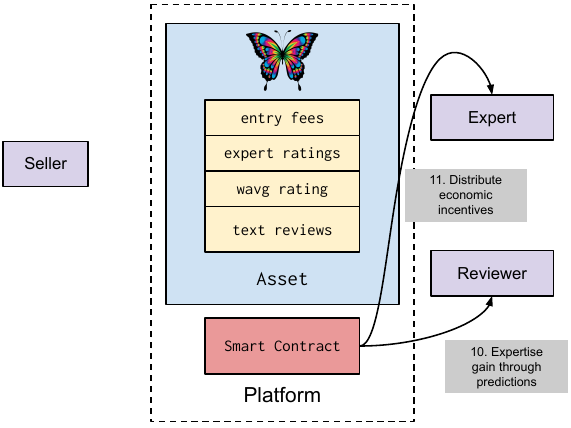}
    \caption{
        Steps involved in the post-sale period.
    }
    \label{fig:postsale}
\end{figure}

\paragraph{Step 10.} \label{arch_prediction}
Once the sale has completed, based on the observed sales metrics, the system calculates a measure of how wrong the ``collective judgement'' of the system was about the popular demand of the asset.
However, the exact way of comparing the popular demand of two assets depends on the sales method used.\footnote[2]{
    For instance, if two assets are sold at predetermined prices without quantity restrictions (e.g. digital copies of games), the gross sales revenue serves as a suitable measure for comparing their sales performance.
    In the case of assets sold through auctions, the final sale prices can be utilized as a metric.
    If the prices are set at fixed amounts and the quantity of assets available for sale is limited, and if the sales were conducted on a first-come, first-served (FCFS) basis, an effective measure would be the time taken for the asset to be sold out.
}
Therefore, we assume for simplicity that the authority chooses some way to obtain the observed popular demand denoted as $demand_a\in[0,1]$ of asset $a$ from some sale metric(s) of its choice.
Furthermore, we also assume that the popular demand predictions made by each reviewer $r$ can be converted in a similar fashion to $prediction_a^r\in[0,1]$ as well.
Then, the individual prediction error of reviewer $r$ on asset $a$ can be calculated as
\begin{equation}
    error(r,a)=(demand_a - prediction_a^r)^2
\end{equation}
and the system-wide prediction error of asset $a$, denoted by $\varepsilon(a)$, can be calculated as
\begin{equation}
    \varepsilon(a) = \frac{\sum_{r\in R} error(r,a) \times exp_r^2}{\sum_{r\in R} exp_r^2}
\end{equation}
where $R$ is the set of all reviewers.
The system-wide prediction error of asset $a$ quantifies the extent to which the review system's collective judgment deviated from the actual popular demand for asset $a$.
This measurement serves as the basis for determining the total amount of expertise to be distributed among the reviewers who participated in making predictions.
The size of the rewards pool increases proportionally with the magnitude of the system-wide prediction error, meaning that the greater its disparity with the true popular demand, the larger the pool of rewards available for distribution among reviewers.
Finally, the rewards pool is distributed among the reviewers with individual shares defined as:
\begin{equation}
    predshare(r,a) =
    \begin{cases}
        0                              & \text{if } error(r,a) \geq \varepsilon(a) \\
        \frac{1}{max(c_2, error(r,a))} & \text{otherwise}
    \end{cases}
\end{equation}
where $c_2$ is a constant used to limit the maximum number of shares any reviewer can obtain for a prediction, determined by the authority.

\paragraph{Step 11.}
Finally, the economic incentive is distributed to the experts that participated in the admission control process, and (optionally) the reviewers that gained expertise through endorsements and/or prediction.
The economic incentive pool can be sourced from any forfeited entry fees from prior rounds, and/or some percentage of the gross revenue from the sales.

\subsubsection{Checks and Balances System}
While having an expert pool with special privileges can provide some protection against spam and malicious entities, it also requires us to actively identify and penalize malicious experts to ensure the proper functioning of the system.
In addition to initial malicious experts, there is a possibility that reviewers may initially act honestly to gain reputation, become experts, and subsequently engage in dishonest behavior.
As the system works on inflationary economics in terms of expertise, incorrect or poor decisions may cause an expert to fall behind others over time, resulting in their removal from the expert pool.
However, this process is slow and not sufficient to penalize all types of malicious actions, such as collusion.

To minimize the impact of malicious experts, we introduce the concept of periodic peer reviews.
During these reviews, a majority vote among the experts can penalize a misbehaving expert by ``burning'' some or all of their expertise, effectively removing them from the expert pool.
The transparency of the blockchain enables the entire history of actions by each expert to be publicly auditable, facilitating this process.
Furthermore, this system of checks and balances also incentivizes experts to endorse other high-quality reviewers.
However, it is important to note that the effectiveness of this checks and balances system depends heavily on the integrity of the initial set of experts.
Therefore, it is crucial to have a majority (at least over $50\%$, preferably higher) of honest experts within the initial set.
We also discuss the effects of having a checks and balances system on the correctness of the architecture in Section~\ref{eval_malicious}.

\section{Evaluation} \label{sec:evaluation}
We analyze the proposed architecture by studying how various design parameters impact the selection of expert reviewers over time.
In particular, we focus on the interplay between two reviewer skill-sets: the ability to subjectively assess the quality of an asset, and the ability to predict popular demand for an asset.
We rely on numerical simulations to study these behaviors.
Our choice of simulations is motivated by the complexity of the system, which makes it challenging to completely model mathematically, and the difficulty of conducting real-world studies on a scale comparable to that of an online NFT marketplace.
Numerical simulations provide us with a way to estimate the system behavior at scale while allowing us to simplify the mathematics involved.

We begin our analysis by considering the simplified scenario in which all participants act \emph{honestly}, i.e, perform actions to the best of their knowledge and abilities.
We draw conclusions about the system behavior based on this scenario before considering the more general case where some participants may act maliciously.
We then determine whether our conclusions hold in the face of such behavior.

\subsection{Simulation with Honest Participants} \label{eval_honest}
\subsubsection{Assumptions}
Each asset $a$ is assumed to have two hidden intrinsic properties\footnote[3]{
    We use two different metrics because critic consensus and popular opinion can often diverge significantly.
    A notable example of a review system employing this concept is Rotten Tomatoes, which displays separate ``Tomatometer'' and Audience scores to capture this disparity.
}
that stay constant throughout the simulation:
\begin{enumerate}
    \item Quality, $q_a \in [0,1]$
    \item Popular Demand, $d_a \in [0,1]$
\end{enumerate}

Similarly, each reviewer $r$ is assumed to have two hidden intrinsic properties that stay constant throughout the simulation:
\begin{enumerate}
    \item Quality Estimation Ability (QEA), $qea_r \in [0,1]$
    \item Popular Demand Prediction Ability (PDPA), $pdpa_r \in [0,1]$
\end{enumerate}

\begin{figure}[htb]
    \centering
    \includegraphics[width=0.5\linewidth]{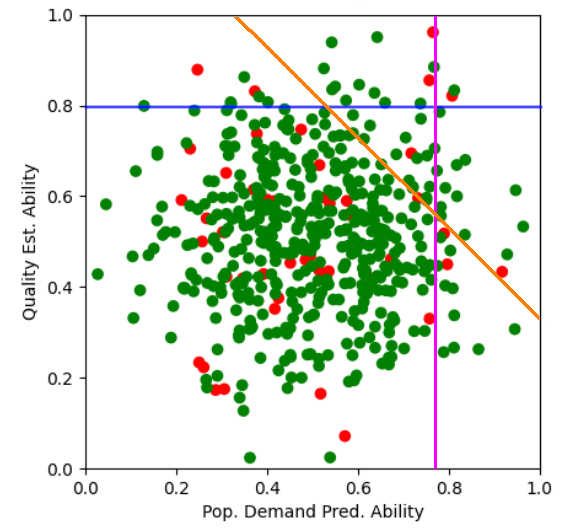}
    \caption{
        A randomly drawn initial population of 500 reviewers, with red dots denoting the initial set of experts.
        The blue, magenta, and orange lines represent slope parameters of $0$, $-\infty$, and $-1$ respectively.
        The ideal final expert set comprises the dots positioned above and/or to the right of the selected slope line.
    }
    \label{fig:initial_slopes}
\end{figure}

Ideally, the authority should have the ability to specify a \emph{slope parameter} within the range of $(-\infty, 0]$.
This parameter determines the relative importance assigned to Quality Estimation Ability (QEA) compared to Popular Demand Prediction Ability (PDPA) when selecting reviewers to become experts.
For instance, as shown in Figure~\ref{fig:initial_slopes}, when the slope is set to $0$, the final set of experts ideally consists of reviewers with the highest QEA.
In contrast, as the slope approaches $-\infty$, the final set of experts should ideally consist of reviewers with the highest PDPA.
By adjusting the slope parameter, the authority can fine-tune the selection criteria for experts based on the desired emphasis between QEA and PDPA.
Throughout the remaining analysis, we assume a slope parameter of $-1$, which assigns equal weightage to both QEA and PDPA.
In this case, the final set of experts ideally comprises the dots located closest to the top right corner of the figure, i.e, points to the right of the diagonal orange line in Figure~\ref{fig:initial_slopes}.

\subsubsection{Simulation}
At the beginning of the simulation, the QEA and PDPA values of each reviewer are randomly sampled from a normal distribution as shown in Figure~\ref{fig:initial_slopes}.
A subset of reviewers is randomly chosen to form the initial expert set and is assigned an expertise value of 100,000, whereas everyone else starts with zero expertise.
We also randomly sample the Quality and Popular Demand of each asset.

Each round in the simulation corresponds to the marketplace life cycle of one asset, as described in Section~\ref{workflow}.
Since we are interested in how the system selects the final expert set, we consider only those products that make it through the admission control process.

We assume that a reviewer's text review of an asset $a$ corresponds to their estimate of $q_a$, and is denoted as $rev_a^r \in [0,1]$.
The magnitude of error of this estimation depends on the QEA of the reviewer.
Concretely, $rev_a^r$ is randomly drawn from the truncated triangular distribution with peak $q_a$ and left and right intercepts determined by $qea_r$ but truncated to the range [0,1].
Thus, the higher the QEA of a reviewer, the more likely $rev_a^r$ will be closer to $q_a$.
Similarly, a reviewer's sales demand prediction of an asset is assumed to be their estimate of $d_a$.
As earlier, $pred_a^r$ is randomly drawn from the truncated triangular distribution with peak $d_a$ and left and right intercepts determined by $pdpa_r$ but truncated to the range [0,1].
It is assumed that all participants of the system produce text reviews and popular demand predictions for all assets.

After the completion of text reviews, we proceed to simulate the endorsement process.
Since we are considering the case where all participants are honest, we assume that each participant will attempt to endorse the review that best aligns with their own assessment.
Concretely, each reviewer $r$ endorses the reviewer $r'$ that minimizes $|rev_a^r-rev_a^{r'}|$.
For each endorsement, we keep track of the expertise gain and investment updates as described in Section~\ref{arch_endorsement}.

Finally, we simulate the sale of the product.
To incorporate market volatility, the sale demand metric is assumed to vary somewhat randomly around the Popular Demand of the asset, simulated using zero-mean Gaussian noise.
Then, we calculate and distribute expertise based on the predictions made by the reviewers as described in Section~\ref{arch_prediction}.
We then move on to the next round of simulation, repeating the process with a new asset.

\subsubsection{Experiments and Results}
\begin{figure}[htb]
    \centering
    \includegraphics[width=\linewidth]{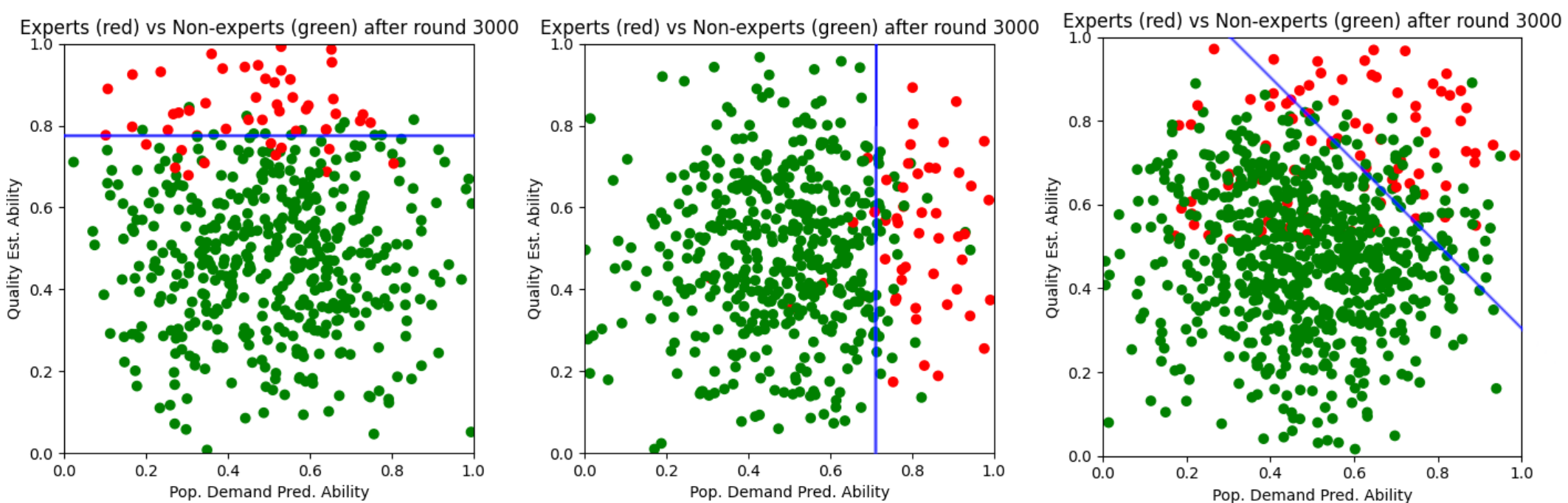}
    \caption{
        Initial and final expert sets with expertise gain enabled from (i) endorsements only (i.e, $slope=0$), (ii) popular demand predictions only (i.e, $slope\approx -\infty)$, and (iii) both sources (i.e, $slope=-1$).
        The convergence behavior was observed even with new reviewers joining the system periodically.
    }
    \label{fig:conv}
\end{figure}

First, we tested the behavior of the system with expertise gain from exactly one of the two sources, i.e, endorsements or popular demand predictions.
This is effectively similar to setting the slope parameter to zero or $-\infty$ respectively.

\begin{figure}[htb]
    \centering
    \includegraphics[width=0.9\linewidth]{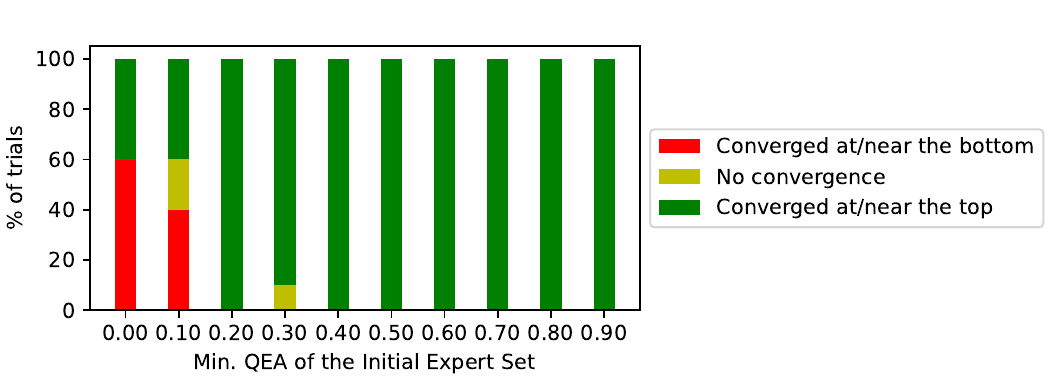}
    \caption{
        Convergence behavior of the system with expertise gain from popular demand prediction set to 0.
        Fig.~\ref{fig:conv}(i) provides an illustrative example of convergence at/near the top.
    }
    \label{fig:qea_vs_conv}
\end{figure}

As seen in Figure~\ref{fig:conv}(i), with the expertise gain from Popular Demand prediction set to zero, the final expert pool after 3000 rounds consists of the reviewers with the highest QEAs.
In particular, we discovered an interesting trend in the simulations: even when the initial set of experts is only ``reasonably'' good in terms of their QEA, with a sufficient number of rounds, the final expert set consists of mostly the reviewers with the highest QEA.
To investigate this phenomenon further, we varied the minimum Quality Estimation Ability of the initial set of experts between 0.1 and 0.9 in steps of 0.1, and repeated the simulation 10 times for each setting.
The results, consolidated in Fig.~\ref{fig:qea_vs_conv}, indicate that the quality of the initial expert list does not necessarily determine the quality of the system.
In other words, past a threshold of rounds, the system ``self-corrects'' by selecting highly skilled experts if all the initial experts have a QEA of at least 0.4.
Similar trends were observed in experiments that allowed expertise gain from popular demand predictions alone, as illustrated in Fig.~\ref{fig:conv}(ii).

Upon enabling expertise gain from both review endorsements and popular demand predictions, it was observed that the final expert pool predominantly consisted of reviewers positioned in the top right corner of the system, indicating high levels of both QEA and PDPA.
This ``convergence phenomenon'' was consistently observed even when new reviewers joined the system over time, as illustrated in Figure~\ref{fig:conv}(iii).
Therefore, based on empirical observations, we conclude that the system is able to select the final expert set by considering a combination of QEA and PDPA, despite these values being hidden from the system.

We then turned our focus to the question of how well the system selects the actual final expert set as compared to the ideal final expert set.
To study this, we systematically varied the minimum QEA of the initial set of experts from 0.1 to 0.9 in increments of 0.1.
For each setting, we repeated the simulation 10 times and recorded the average \emph{combined score} of the actual and ideal final expert sets.
In this context, the combined score is obtained by taking the mean of the QEA and PDPA values, as a slope parameter of $-1$ gives both factors equal importance.
The summarized results are presented in Table~\ref{tab:convergence}.

\begin{table}[htb]
    \centering
    \begin{tabular}{|l|c|c|c|c|c|c|c|c|c|c|}
        \hline
        \textbf{Min. Quality Est. Ability $\rightarrow$} & \textbf{0.0} & \textbf{0.1} & \textbf{0.2} & \textbf{0.3} & \textbf{0.4} & \textbf{0.5} & \textbf{0.6} & \textbf{0.7} & \textbf{0.8} & \textbf{0.9} \\
        \hline
        Initial Expert Set                               & 0.50         & 0.48         & 0.47         & 0.58         & 0.57         & 0.59         & 0.6          & 0.64         & 0.72         & 0.72         \\
        \hline
        Ideal Final Expert Set                           & 0.76         & 0.77         & 0.73         & 0.76         & 0.78         & 0.78         & 0.77         & 0.76         & 0.78         & 0.80         \\
        \hline
        Actual Final Expert Set                          & 0.57         & 0.61         & 0.58         & 0.64         & 0.72         & 0.71         & 0.72         & 0.72         & 0.72         & 0.70         \\
        \hline
    \end{tabular}
    \caption{Mean combined score ($=\frac{QEA+PDPA}{2}$) of Actual vs Ideal (best 50) final expert set, compared to the initial expert set.}
    \label{tab:convergence}
\end{table}

Table~\ref{tab:convergence} reveals an interesting trend.
When the initial set of experts has a minimum QEA of 0.3 or lower, the resulting final expert set demonstrates a significantly lower mean combined score compared to the ideal final expert set.
Similarly, at the other extreme, when the initial set of experts has a minimum QEA of 0.8 or higher, the resulting final expert set performs at the same level as, or sometimes even worse than, the initial expert set.
Between the two extremes, we identify a ``sweet spot'' for the minimum QEA of the initial expert set within the range $[0.4, 0.6]$.
In this range, the actual final expert sets exhibit substantially higher quality than the initial set.
Furthermore, the mean combined score of the actual final expert set is consistently close to that of the ideal final expert set.

\subsection{Simulation with Potentially Malicious Participants} \label{eval_malicious}
In the previous section, we made the assumption that all participants in the system would act honestly.
However, in real-world deployments, this is never the case.
To comprehensively examine the risks posed by malicious participants, we develop a threat model that considers a wide range of different malicious actions that participants could perform.
We then analyze the negative impact each malicious action can have on the system, the incentives driving such behavior, the safeguards in place to discourage such actions, and the scope and potential consequences of each malicious action based on the incentives and safeguards identified.
The exhaustive list of malicious actions that can be performed by the participants of the proposed review system are as follows:

\paragraph{1. An expert provides a dishonest rating during admission control.}
\begin{itemize}
    \item \textbf{Adverse effects:}
          High-quality assets may be rejected from the marketplace, while low-quality assets may be accepted.

    \item \textbf{Incentives for malicious behavior:}
          Off-platform incentives, such as a bribe from the seller.

    \item \textbf{Safeguards:}
          Frequently misbehaving experts can be identified and penalized through periodic internal reviews.
          Furthermore, it is in the interest of experts to only let the best products through admission control because of two reasons.
          Firstly, the economic incentive distributed to the experts depends on the sales amount.
          Secondly, the size of the popular demand prediction rewards pool depends on how wrong the expert consensus was.

    \item \textbf{Evaluation:}
          First, we point out that off-platform incentives are difficult to manage through on-platform interventions in any decentralized system, and our proposed system is no different.
          However, we argue that the safeguards offer sufficient protection:
          It may not be economically viable for the seller to pay a large bribe to a majority of experts if the product is of low quality (i.e, likely to have weak sales).
          Thus, the larger the expert pool, the harder it gets for the seller to bribe a majority of experts.
          Furthermore, with frequent internal reviews, misbehaving experts can be identified and penalized.
\end{itemize}

\paragraph{2. An expert acts maliciously during the periodic internal review.}
\begin{itemize}
    \item \textbf{Adverse effects:}
          Honest experts may be incorrectly penalized.

    \item \textbf{Incentives for malicious behavior:}
          A group of malicious experts might attempt to remove ideologically different experts from the pool to consolidate power.
          Off-platform incentives may also be a driving factor.

    \item \textbf{Safeguards:}
          The periodic internal review is self-correcting, i.e, if most experts are honest, then the experts that act dishonestly during periodic internal reviews will be penalized.

    \item \textbf{Evaluation:}
          We argue that the periodic internal review provides sufficient safeguards as long as most experts are honest.
\end{itemize}

\paragraph{3. A reviewer provides a dishonest text review / popular demand prediction for a product (after it has been listed on the marketplace)}
\begin{itemize}
    \item \textbf{Adverse effects:}
          Dishonest text reviews \emph{may} have some influence on the public opinion of a product, but their impact is limited as reviews are ranked based on the reviewer's expertise.

    \item \textbf{Incentives for malicious behavior:}
          Off-platform incentives.

    \item \textbf{Safeguards:}
          Reviewers have no incentive to provide dishonest popular demand predictions, as it could result in them losing popular demand prediction rewards.
          Moreover, assuming most experts act honestly, the likelihood of a dishonest text review receiving expert endorsements is low.

    \item \textbf{Evaluation:}
          The adverse effects of a dishonest text review are limited, and does not influence the admission control decision in any way.
\end{itemize}

\paragraph{4. An expert or reviewer dishonestly endorses a particular text review.}
\begin{itemize}
    \item \textbf{Adverse effects:}
          Dishonest endrosements by experts can lead to undeserving individuals gaining expertise.
          Over time, this may result in an expert pool consisting of low-quality or dishonest reviewers, jeopardizing the integrity of the system.

    \item \textbf{Incentives for malicious behavior:}
          Experts may have an incentive to endorse ``bad'' reviewers to prevent them from becoming experts and potentially displacing existing experts in the pool.
          Non-expert reviewers, however, do not have similar incentives for making dishonest endorsements.

    \item \textbf{Safeguards:}
          To address the risk of experts endorsing bad reviewers, we introduce the ``investment'' concept.
          This allows experts/reviewers to earn a share of the future expertise earned by those they endorse.
          However, further assessment is required to determine the effectiveness of this safeguard.
          
    \item \textbf{Evaluation:}
          We need to evaluate different endorsement strategies by experts to determine whether it is in their interest to act honestly.
\end{itemize}

Based on the above analysis, we concluded that the most significant malicious action deserving in-depth study is the act of reviewers engaging in selfish endorsements.
This pertains to expert or non-expert reviewers endorsing text reviews for motives other than genuine alignment with their own views.
Specifically, we focus on the following two questions:
\begin{enumerate}
    \item Does adopting a selfish endorsement strategy provide long-term benefits for experts compared to utilizing an honest endorsement strategy?
    \item How does the endorsement strategy employed by non-expert reviewers affect the convergence of the system?
\end{enumerate}

\subsubsection{Selfish Endorsement Strategies}
We start by defining the following selfish endorsement strategies:
\begin{enumerate}
    \item \emph{Lazy Endorsement.} This strategy involves endorsing a randomly chosen text review without considering its quality.
    \item \emph{Endorse Another Expert.} Under this strategy, the reviewer intentionally endorses another expert to either maintain the \emph{status quo} within the expert pool or potentially increase their own future investment dividends.
    \item \emph{Endorse a Poor Reviewer.} In this strategy, the reviewer purposely endorses a reviewer who exhibits low quality or competence, with the intention of preventing them from accumulating enough expertise to challenge the status quo within the expert pool.
    \item \emph{No Endorsement.} This strategy involves refraining from endorsing any review, thereby denying any reviewer the opportunity to gain expertise through the endorsement.
\end{enumerate}

\begin{figure}[htb]
    \centering
    \includegraphics[width=0.85\linewidth]{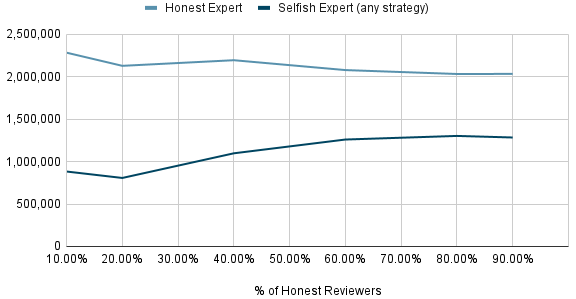}
    \caption{
        Final average expertise by strategy.
        For each experiment, we selected the selfish strategy that yielded the highest average final expertise and plotted it alongside the average final expertise of the honest experts.
    }
    \label{fig:strategies}
\end{figure}

\subsubsection{Experiments and Results}
We conducted a series of experiments by varying the percentage of honest experts from 10\% to 90\%.
The remaining experts were assigned different combinations of the four selfish strategies outlined previously.
For each experiment, we recorded the average final expertise of the honest experts and experts employing any of the four selfish strategies.
These results are consolidated in Figure~\ref{fig:strategies}.
From the figure, it is evident that the honest strategy consistently outperforms every analyzed selfish strategy in the long run.
Additionally, the other selfish endorsement strategies only offer marginal improvements over not endorsing any review at all.
Based on these empirical findings, we can now answers the questions posed earlier.

Firstly, the analysis of the four selfish endorsement strategies reveals that none of them offer any long-term benefits to the experts.
On the contrary, the simulations clearly show that experts who endorse honestly have a significant advantage over those employ selfish endorsement strategies.
Secondly, similar to experts, non-expert reviewers who employ an honest endorsement strategy were observed to outperform those who adopted selfish strategies.

Based on our analysis, we can conclude that the investment concept serves as an effective safeguard against reviewers engaging in selfish endorsements, which is the primary malicious action that participants can undertake within the system.
Therefore, the results obtained in Section~\ref{eval_honest}, which were based on the assumption of honest participants, can be applied more broadly to scenarios where the majority of experts are not malicious.
It is worth noting that other potential malicious actions that have not been considered in our threat model may require further investigation and safeguards.
Nonetheless, our findings support the conclusion that the investment concept, combined with honest participation, establishes a robust foundation for an effective and reliable review system.

\section{Conclusion} \label{sec:conclusion}
We have introduced a decentralized review system specifically designed for marketplaces that deal with highly scarce and highly resellable products, particularly focusing on NFT marketplaces.
However, it is important to note that the proposed review system can be applied to any marketplace that involves the sale of extremely scarce products with significant resale potential.
This can include various assets typically auctioned at specialized platforms or auction houses.
The fundamental principles and mechanisms of our review system can be adapted and tailored to suit the specific characteristics and dynamics of different marketplaces, ensuring transparency, credibility, and reliability in evaluating and assessing the products being traded.

Future work could explore the implementation and deployment of such a system on a practical platform and evaluate with real users. Extensions to domains beyond art or game-related NFT marketplaces may also be of interest.

\section*{Acknowledgements}
This research was funded by the Sony Research Award Program.
This paper has been edited with the assistance of ChatGPT.
We certify that ChatGPT was not utilized to produce any technical content, and we accept full responsibility for the contents of the paper.

\bibliographystyle{splncs03}
\bibliography{references}

\end{document}